# Rejoinder: The 2005 Neyman Lecture: Dynamic Indeterminism in Science

**David R. Brillinger**

I was so fortunate as to spend formative periods of my statistical career watching and working near two of the powerhouses of twentieth-century statistics—Jerzy Neyman (JN) and John Tukey. The first championed the responsibility the statistician has to set down a clear pertinent set of assumptions guiding her/his data analyses. The second emphasized the importance of looking for discoveries and surprises in data sets.

The Neyman Lecture gave me an opportunity to show my admiration for Professor Neyman and his applied work. The examples from my own work are meant to parallel analyses from his work. In some cases the analyses were done some years ago. The paper may be considered a substantial update of Brillinger (1983). Both Grace Yang and Hans Künsch add meat to the paper and thereby increase our understanding of Jerzy Neyman and his contributions.

I begin with Grace's Discussion. Her comments "resonate" with me, to use her word. Indeed her Discussion, with its emphasis on Neyman's teaching and research projects on sampling and cancer, creates here a collaborative paper concerning Neyman's applied statistics career.

As well as lively anecdotes, Grace presents some Neyman quotes. One that she found that I like particularly is,

> I deeply regret the not infrequent emphatic declarations for or against pure theory and for or against work in applications. It is my strong belief that both are important and, certainly, both are interesting.

*David Brillinger is Professor, Department of Statistics, University of California, Berkeley, California 94720, USA (e-mail: brill@stat.berkeley.edu).*

The various quotes plus Grace's own words bring out Neyman's approach to science in general and statistics in particular. I refer you to the second paragraph in her section "Neyman as a teacher and his problem-driven approach." Grace further emphasizes today's appearance of massive data sets and the steady appearance of data of novel types that may be perceived as realizations of stochastic processes. She focuses on the Neyman–Fix competing risks model and on a Markov-branching model for the effect of radiation. There are figures displaying yeast cell survival data and the results of fitting a science-based model.

Grace refers to the importance of point processes. I mention in admiration that I regard Yang (1968) as one of the earliest statistics papers bringing a nontrivial point process analysis into a statistical analysis. Her expression (2.1) in the 1968 paper in a sense introduces the conditional intensity function, a concept that has proved an incredibly powerful tool in both theory and applications.

Hans Künsch's Comments are of a different character, and give me an opportunity to elaborate on some of the material in the paper and to mention my future directions. Also let me say that, like Grace's Discussion, I do not find anything in Hans's that I disagree with. Hans chides my analyses some, and then leads the reader into the modern world of simulation, and stochastic difference equation (SDE) methods. (Let me remark somewhat defensively that every scientific paper is a progress report and apologize for not having provided enough detail in some cases.)

Concerning SDEs, Hans mentions the lack of bounded variation of their paths and the natural unreasonableness of this. (This provides an explanation of why one can estimate the parameter $\sigma$ with probability 1, by the way.) I saw Brownian-based SDEs as a convenient motivator for stochastic models of trajectories and consequent data analyses. Their uses include provision of convenient approximations to Markov processes in discrete time. As generally formulated, however, they lead to Markov processes, which animal tracks are not, for animals eat and then there is a period when they do not eat, for example. To handle this I am now including time lags in the Brownian SDE model, leading to non-Markov processes. I am also working with noise processes other than the Brownian, and consequently the Stratonovich form of SDEs. This allows inclusion of general lagged time effects.

Now some remarks concerning the rainfall example. The data for Figure 4 were derived from a figure in Neyman and Scott (1974) by reading off values. Each curve involves but 24 values and the points joined are 3-point moving averages of hourly values. The individual day values as well as the hourly averages appear to be lost. This reduced data set seems to mean that Hans's proposed random effect model suffers from aliasing of the within and between day effects. Hans calls for more science in the modeling of the rainfall. Indeed. That is certainly in the Neyman spirit, as emphasized by Grace. Some meteorology and wind analysis had been included by JN and his collaborators by virtue of the days chosen having been selected to have winds from the south and a warm stability layer. The suggestion of a decaying intensity does seem reasonable. Lastly, I certainly bow to Hans in the matter of knowledge of rainfall in Zurich.

Next the blowflies. The results presented are from 1978. The need for a "complete" model is clear and mentioned in Brillinger et al. (1980). Guttorp (1980) and I began work which we have not yet gotten back to. One contribution of the 1980 paper is an early use of the Kalman filter in applied statistics. The state-space model is particularly appropriate for work with age-structured populations.

Lastly the seals. In further studies of the elephant seal tracks (Brillinger, 2000), simulation is employed to estimate the transition density of the track process in the presence of measurement error and then maximum likelihood estimates computed from an approximate likelihood. Continuing, working with an approximation like (3) appeared simpler generally than working with a transition density, that is, the Fokker–Planck approach. Indeed the model for the data was taken to be (3) and, for example, examined via residual plots. More details may be found in Brillinger et al. (2001). Hans mentions the paper Jonsen, Mills, Flemming and Myers (2005). One of its important contributions is the necessary provision of robust/resistant estimates of animals' tracks from scattered locations. (I have to mention, though, that personally I prefer working with the unequally spaced values as long as possible rather than interpolating.)

More of the JN story may be found in the recent book of Lehmann (2008).

I thank, as I am sure do many readers, Grace and Hans for adding, expanding and clarifying the topics of this paper and for presenting pertinent new material.